\documentclass[preprint]{aastex}

\begin{document}


\title{A Near-Infrared Imaging Survey of Coalsack Globule 2}


\author{Germ\'an Racca\altaffilmark{1}, Mercedes G\'omez\altaffilmark{2}}
\affil{Observatorio Astron\'omico de C\'ordoba, Laprida 854, 5000 C\'ordoba,
Argentina}
\email{german@oac.uncor.edu, mercedes@oac.uncor.edu}      

\and

\author{ Scott J. Kenyon\altaffilmark{2}}
\affil{Smithsonian Astrophysical Observatory, 60 Garden Street, Cambridge MA 02138}
\email{skenyon@cfa.harvard.edu}


\altaffiltext{1}{Also at the Facultad de Matem\'atica, Astronom\'\i a y F\'\i sica, 
Universidad Nacional de C\'ordoba, Argentina.} 

\altaffiltext{2}{Visiting Astronomer, Cerro Tololo Inter-American Observatory.
CTIO is operated by AURA, Inc.\ under contract to the National Science
Foundation.}


\begin{abstract}

We describe a near-infrared imaging survey of Globule 2 in the Coalsack.  
This Bok globule is the highest density region of this southern hemisphere 
molecular cloud and is the most likely location for young stars in 
this complex.  The survey is complete for $K$ $<$ 14.0, $H$ $<$ 14.5, 
and $J$ $<$ 15.5, several magnitudes more sensitive than previous 
observations of this globule.  From the large number of background stars,
we derive an accurate near-infrared extinction law for the cloud. Our result,
$E_{J-H}/E_{H-K}$ = $2.08 \pm 0.03$, is significantly steeper than results 
for other southern clouds. We use the $J-H$/$H-K$ color-color diagram to 
identify two potential young stars with $K$ $<$ 14.0 in the region.
We apply $H$-band star counts to derive the density profile of the Coalsack
Globule 2 and use a polytropic model to describe the internal structure of this small cloud. 
For a gas temperature T $\sim$ 15 K, this globule is moderately unstable.  
 
\end{abstract}


\keywords{ISM: individual (Coalsack) --- ISM: dust, extinction ---
ISM: globules --- star: formation --- star: pre-main sequence}


\section{Introduction}

The Coalsack is a conspicuous, nearby \citep[d $\sim$ 180 pc;][]{fra89}
dark cloud located close to the Galactic plane in the southern Milky Way
(l $\sim$ 303$^o$, b  $\sim$ 0$^o$). \citet{nym89} estimated a total mass of 
$\sim$ 3550 M$_{\odot}$ from $^{12}$CO data covering an area of $\sim$ 15 deg$^2$
in the direction of this complex. Star counts toward the same area indicate 
an average optical extinction of $A_V \sim$ 5 mag \citep{gre88,cam99}. 
However, the extinction is $A_V \sim$ 20 mag or larger in several small dark globules
\citep{tap73,har86,bou95a}.  This structure is similar to other well-known 
clouds such as Taurus, Chamaeleon I, and Lupus that are actively forming stars
\citep{ceba84,bou98,argo99,cam99}.  Despite this similarity, previous searches
of the Coalsack have failed to detect signs of recent star-formation 
\citep{wea74,sch77,rei81,nym89}.

Recently \citet{kat99} have observed the Coalsack in the $^{13}$CO and 
C$^{18}$O J$=$1--0 lines. The $^{13}$CO data show a massive cloudlet 
($\sim$ 200 M$_{\odot}$) located at the western edge of the complex, 
coinciding with the region of the largest optical extinction 
\citep[see][]{cam99}. In addition, these authors identified five 
C$^{18}$O cores \citep[see also][]{vil94}, corresponding to the positions 
of well-known optically dark globules \citep{tap73, har86, bou95a}.  
These cores have high column densities, typical of star-forming cores, 
and masses between 4--10 M$_{\odot}$. However, there are no {\it IRAS}
sources, and thus no evidence of recent star-formation, in these cores
\citep{kat99}.

Tapia's Globule 2 ($\alpha$ $=$ 12$^h$ 31.5$^m$, $\delta$ $=$
$-$63$^o$ 44.5$'$; 2000.0) is the densest 
(n(H$_2$) $\sim$ 4 $\times$ 10$^3$ cm$^{-3}$) and most massive 
($\sim$ 10 M$_{\odot}$) of the Kato et al. cores 
\citep[see also][]{bou95b}.  It is an obvious, roughly circular 
patch \citep[$\sim$ 6$'$ radius;][]{bou95a} of extinction on 
shallow optical and near-infrared (near-IR) surveys. \citet{jon80} 
observed an area of $\sim$ 850 arcmin$^2$ to $K=$ 9.5; \citet{jon84} 
scanned an area of 2$'$ $\times$ 2$'$, 70\% complete to $K=$ 13.7,
with the central 1$'$ $\times$ 1$'$ region 70\% complete to $K=$ 14.7. 
Neither of these surveys detected candidate near-IR excess stars.
At $K=$ 9.5, \citet{jon80, jon84} could detect $\sim$ 1 M$_{\odot}$ main 
sequence stars at the distance of the Coalsack.  In the central 
1$'$ $\times$ 1$'$ region, 70\% complete to $K=$ 14.7, this detection 
limit is $\sim$ 0.4 M$_{\odot}$, assuming A$_K$ $\sim$ 2 
\citep[see][]{hemc93,del00}. 

The Coalsack Globule 2 has also been observed by the {\it Midcourse Space
Experiment}  \citep[MSX;][]{pri01}. Fits format images and a source catalog
in four mid-infrared bands, A(8.28 $\mu$m), C(12.13 $\mu$m), D(14.65 $\mu$m),
and E(21.34 $\mu$m),  are available through the NASA/IPAC IR Science
Archive (IRSA)\footnote{http://irsa.ipac.caltech.edu.}.
Nine mid-infrared sources lie within \hbox{$\sim$ 11$'$} from the optical
center of the globule. One of these is located $\sim$ 2.8$'$ SW 
from the center of the most massive Kato et al. C$^{18}$O cores.
This different is comparable to the 2$'$ resolution of the
molecular line maps. 

Here, we describe a near-IR imaging survey of $\sim$ 
15$'$$\times$15$'$ centered on Globule 2. The observations are 
uniformly complete to $K=$ 14.0, $H=$ 14.5, and $J=$ 15.5.  
If $A_K \sim$ 0--2 mag, these data are sensitive to $\sim$ 0.2--0.5 M$_{\odot}$ 
main sequence stars \citep[see][]{hemc93,del00}.  
We detect $\sim$ 6500 sources to the detection limits 
of $K=$ 16.5, $H=$ 17.0, and $J=$ 18.0.  We restrict our analysis to 
$K=$ 14.0 ($\sim$ 2500 sources) where the photometric errors are 
relatively small ($<$ 0.08 mag). 

In \S 2 we describe the observations and data reduction. 
In \S 3 we derive a reliable extinction curve for background stars 
in the Globule 2 region and use this result in \S 4 to search for potential 
young stellar objects in our survey region based on their locations
in the $J-H$/$H-K$ diagram. We identify only two objects with 
$K$ $<$ 14.0 and near-IR excesses characteristic of young stellar 
objects. In \S 5 we use $H$-band stars counts to derive the 
density profile of this small cloud and compare it with the
predictions of polytropic models. Assuming T $\sim$ 15 K, 
Globule 2 is moderately unstable.  We conclude with a
brief summary in \S 6.

\section{Observations and Data Analysis}

We obtained $JHK$ imaging data for Globule 2 in the Coalsack and four 
relatively unreddened control fields on 13--16 February 1995, and
8--11 March 1996, with CIRIM (the Cerro Tololo Infrared Imager) at 
the CTIO 1.5m telescope. The CIRIM uses a 256 $\times$ 256 HgCdTe NICMOS 3 
array, which provides a field of $\sim$ 4.9$'$ $\times$ 4.9$'$ with
a plate scale of 1.16$''$ per pixel. We covered an area of 
\hbox{$\sim$ 15$'$ $\times$ 15$'$} on the cloud on a regular grid,
with 1$'$ overlap between adjacent frames. We acquired two 6 $\times$ 5 
sec exposures for each field, shifted by 20$''$. We also observed
four 5$'$ $\times$ 5$'$ control fields, selected from a visual inspection
of the ESO Red Sky Survey prints, close to Globule 2 region and relatively
free from significant optical extinction. Table 1 lists positions of these 
off-cloud regions.

We process the data using standard techniques with the
IRAF\footnote{IRAF is distributed by the National Optical 
Astronomy Observatory, which is operated by the Association of 
Universities for Research in Astronomy, Inc. under contract to the
National Science Foundation.} software package. \citet{goke01} 
describes the data
reduction procedure in detail.  Briefly, we apply dark, flatfield, 
and sky corrections to all images, align dithered pairs using the 
IRAF subroutines GEOMAP and GEOTRAN, and then add the dithered pairs.
We select the $K$ images as the reference frames and transform the $J$ 
and $H$ images to the same pixel scale as the $K$ images. We use DAOFIND 
to locate stars 4$\sigma$ above the local background and add to the 
DAOFIND list all stellar objects missed by this routine and found by visual
inspection of each image. We then derive photometry for each image using
the APPHOT PHOT task, using a circular aperture with 5$''$ radius. 
We detect $\sim$ 6500 sources at $JHK$ in our survey region. 

To calibrate our photometry, we observed on each night a set of 10-15 
standards from \citet{eli82} and from the UKIRT faint $JHK$ standard 
stars list \citep{haw01}. \citet{goke01} estimate photometric uncertainties
of $\pm$ 0.03 mag in this calibration. These authors also list typical
photometric uncertainties in each magnitude bin. These uncertainties
reach 0.3 mag for $J$ $\sim$ 17, $H$ $\sim$ 16, and $K$ $\sim$ 15.
The 5$\sigma$ limiting magnitudes are $K$ $=$ 14.5, $H$ $=$ 15.5, and
$J$ $=$ 16.0. The magnitudes are in the CIT \citep{eli82} system 
\citep[see][]{goke01}.
                                       
To check the completeness of the survey, we construct a list of detected 
sources. Figure 1 shows the magnitude histograms of background stars as 
functions of $JHK$. The number of stars per bin increases monotonically 
up to the completeness limit and then turns over.  The corresponding 
limits in each filter are $J$ = 15.5, $H$ = 14.5, and $K$ = 14.0.

\begin{figure}
\centering
\includegraphics[width=17cm]{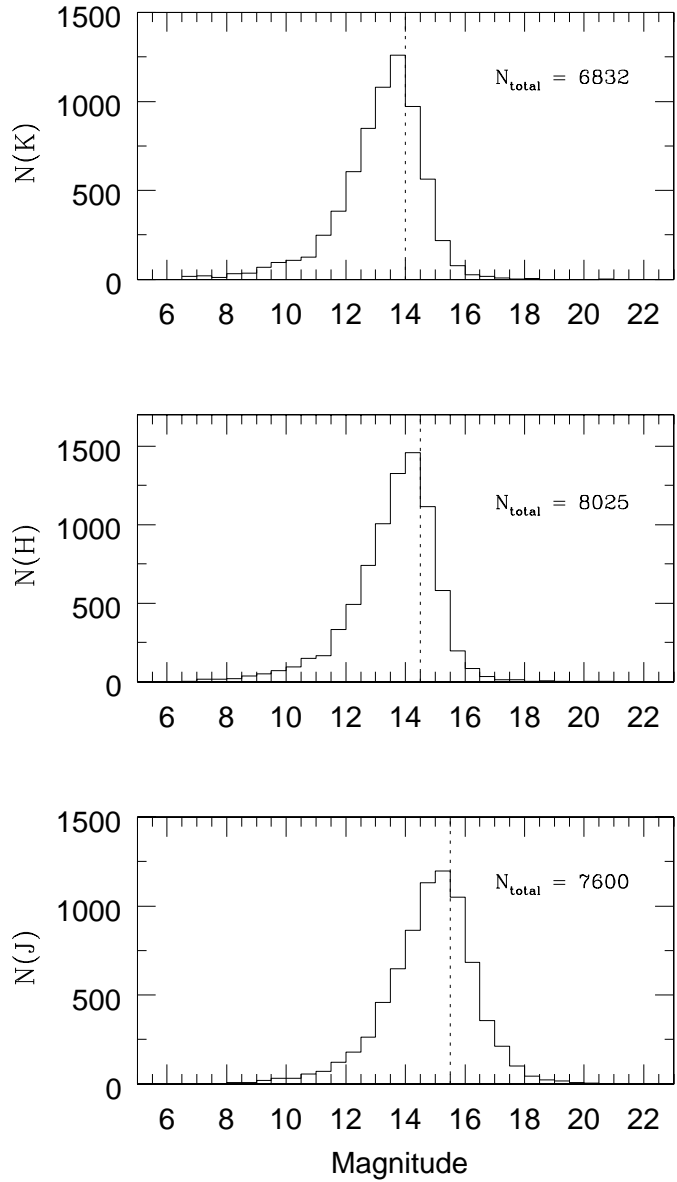}
\caption{Histogram distribution of magnitudes for background stars
in our survey region.  The completeness limit in each band is indicated by the
dashed vertical lines.}
\label{Fig1}
\end{figure}         

To derive coordinates for our survey stars, we use WCSTools\footnote{
Available at ftp://cfa-ftp.harvard.edu/pub/gsc/WCSTools} 
\citep{min97}, a suite of programs to calculate a direct transformation 
between the coordinates of the image (x,y) and the sky coordinates 
($\alpha$,$\delta$). We measure the transformation coefficients adopting
matches between program stars and stars in the U.S.  Naval Observatory
SA1.O Catalogue \citep{mon96}. We estimate an average uncertainty in our
positions of 1$''$.  This estimate relies on a direct comparison of our 
coordinates with coordinates from the  Digitized Sky
Survey\footnote{Based on photographic data obtained using The UK Schmidt
Telescope. The UK Schmidt Telescope was operated by the Royal Observatory
Edinburgh, with funding from the UK Science and Engineering Research
Council, until 1988 June, and thereafter by the Anglo-Australian
Observatory. Original plate material is copyright (c) the Royal
Observatory Edinburgh and the Anglo-Australian Observatory. The
plates were processed into the present compressed digital form
with their permission. The Digitized Sky Survey was produced at
the Space Telescope Science Institute under US Government grant NAG W-2166.
Copyright (c) 1993, 1994, Association of Universities of Research in
Astronomy, Inc. All right reserved.} (DSS). 




\section{The Coalsack Globule 2 Near-Infrared Reddening Law}

We use a general photometric technique to derive the near-IR reddening
law in the Coalsack \citep{ke98}.  The technique assumes that the 
stellar population behind the cloud is identical to the stellar
population in off-fields several degrees away.  Because the Coalsack
contains so few pre-main sequence stars, this assumption is reasonable.
\citet{ke98} derive the $J-H$ and $H-K$ color excesses for each on-cloud 
source relative to every off-cloud source and then compute the average 
and median color excesses for each on-cloud source.  The probable error 
of the average color excess is the sum in quadrature of the errors of 
the on-cloud and off-cloud colors. For the median color excess, the 
probable error is the inter-quartile range.  

Figure 2 shows average color excesses of Coalsack stars with K $\le$ 14.
We consider stars with \hbox{$K$ $<$ 14.0}, where the photometric errors
are $\lesssim$ 0.1 mag.  We eliminate two potential young stars with
near-IR excesses from this analysis (see \S 4 below).  The color excesses 
are correlated, with a Spearman rank-order
correlation coefficient of $r_s$ $=$ 0.88. The probability for no 
correlation between the two color excesses is formally zero according 
to the Spearman rank-order test. Straight line fits to color excess 
measurements yield $E_{J-H}/E_{H-K}$ $=$ 2.07 $\pm$ 0.05.  Median color 
excesses for each source yield the same slope and error.

\begin{figure}
\centering
\includegraphics[width=17cm]{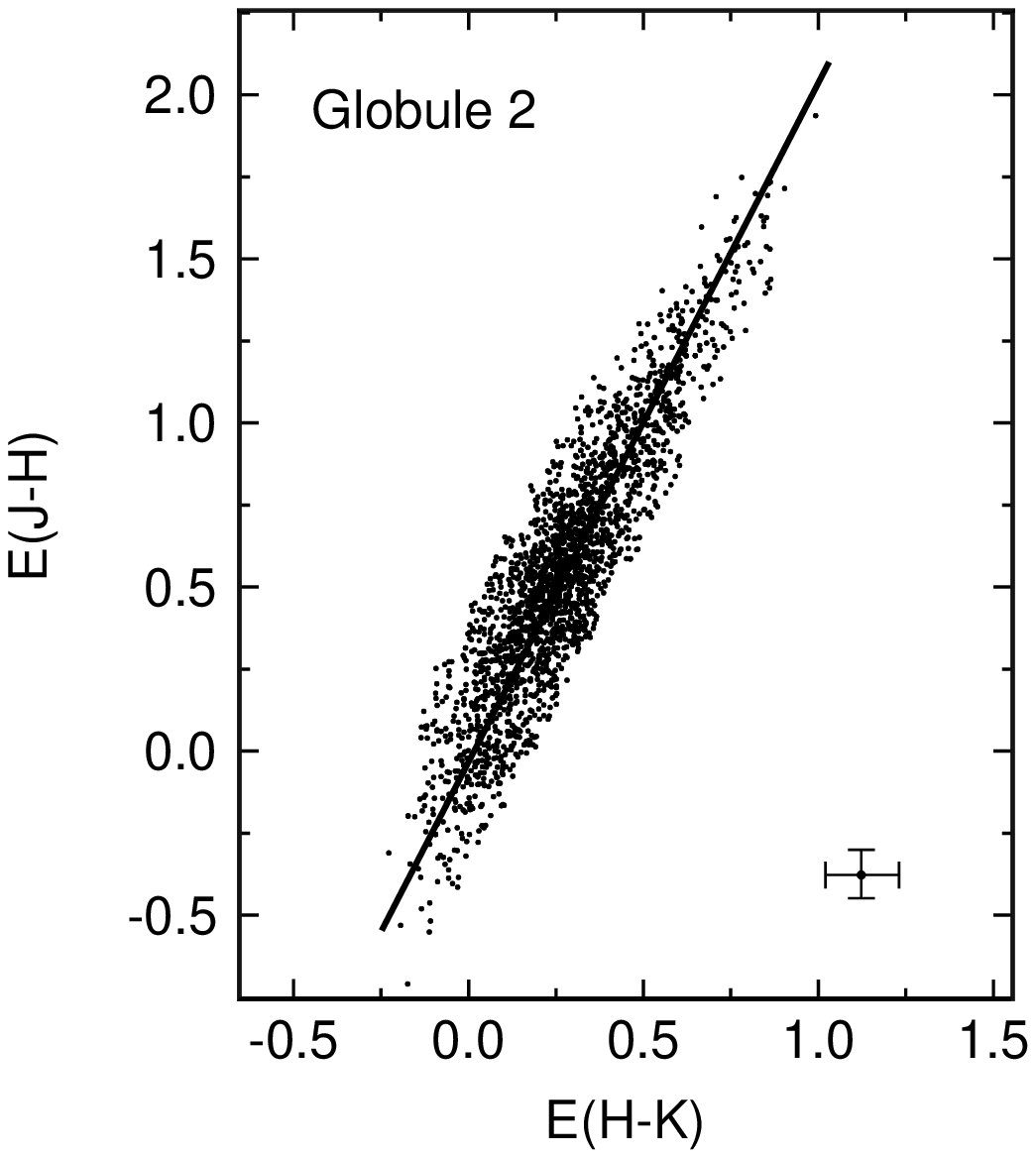}
\caption{Average color-excess diagram for Globule 2
stars with $K < 14$.  The sample includes 2271 stars used in the
analysis.}
\label{Fig2}
\end{figure}         

To test the robustness of the near-IR extinction law, \cite{ke98}
define the reddening probability function.  The reddening probability
$\rho(E_x,E_y)$ is the chance of measuring a pair of color excesses 
$E_x$ and $E_y$, where $x$ and $y$ are color indices and
$N = \int \rho(E_x,E_y) ~ dE_x dE_y$ is the number of reddening
measurements. The number, $N$, is also the number of on-field stars
in the sample.  The density function, $\rho(E_x,E_y)$, depends on
the distributions of colors in the on-field and the off-field
and results from a convolution of the densities of off-field
and on-field stars in the color-magnitude diagram \citep{ke98}.
These color distributions use a kernel density estimator with smoothing
parameter $h$ = 0.2 to derive the density.

Figure 3 shows the probability functions for the Coalsack.  The color-color
diagram in the left panel shows the middle of the main sequence and the
lower main sequence as peaks in the density.  The middle panel indicates
a clear peak in the sample of Coalsack stars at $H-K \approx$ 0.5 and
$J-H \approx$ 1.3--1.5.  The reddening probability function in the right
panel consists of a collection of nested aligned ellipses.  The major
axes of these ellipses have a slope consistent with the slope of
the reddening law in Figure 2, $E_{J-H}/E_{H-K}$ $=$ 2.08 $\pm$ 0.03.
\citet{ke98} describes the derivation of the slope of the major axis 
of each contour in Figure 3.  The ellipses are symmetric about the 
reddening line, suggesting that the Coalsack contains few stars with
near-IR excesses (see below).

\begin{figure}
\centering
\includegraphics[width=17cm]{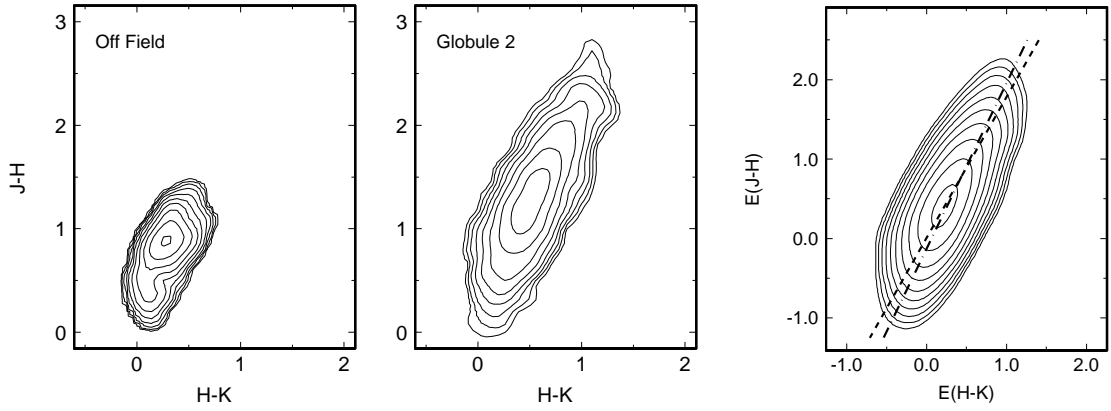}
\caption{Density distributions for Globule 2 stars.
The left and middle panels show the color density distributions for
stars off the main Globule 2 field (off-field, left panel) and for
stars projected on the globule (Globule 2, middle panel).  The right
panel plots the reddening probability distribution for the Globule 2
sources, assuming that these stars have the same color distribution
as the off-field stars.}
\label{Fig3}
\end{figure}        

The Coalsack near-IR extinction law is different from other near-IR 
extinction laws.  In previous studies, we derive
$E_{J-H}/E_{H-K}$ $=$ 1.80 $\pm$ 0.03 for the Chamaeleon I dark cloud 
\citep{goke01} and
$E_{J-H}/E_{H-K}$ $=$ 1.57 $\pm$ 0.03 for the $\rho$ Ophiuchi dark cloud
\citep{ke98}.
The \hbox{Cha I} extinction law agrees with the standard extinction law of 
\citet{bebr88}, $E_{J-H}/E_{H-K}$ $=$ 1.75, transformed to the CIT
system.  The $\rho$ Oph law is more similar to results from the
\citet{he95} survey of luminous southern stars, $E_{J-H}/E_{H-K}$
$=$ 1.47 $\pm$ 0.06.  The extinction laws for the three dark clouds
differ by $\sim$ 10$\sigma$, suggesting dramatic changes in
grain composition or size between these clouds.

The ratio of total to selective extinction, R$_V$, may also to
vary among dark clouds \citep[see][and the references therein]{ke98}.
\citet{kima96} attribute this variation to different
grain properties from cloud to cloud. Dense regions have larger
R$_V$ and fewer small grains than diffuse clouds. On average, grain
sizes increase with cloud density. In addition, recent calculations
show a clear dependence of the extinction law on the grain mixture
and composition \citep[see, for example,][]{vai01}.    

Molecular line observations confirm the difference between Cha I and
the Coalsack.  \citet{har96} report a significant range in the
ratio of the CO column density $N(CO)$ to the $J-K$ color excess
$E_{J-K}$ in several southern molecular clouds.  The observed $N(CO)/E_{J-K}$
is a factor of two larger in Cha I than in the Coalsack.  The R~CrA 
star-forming region also has a large $N(CO)/E_{J-K}$.  \citet{har96}
suggest that $N(CO)/N(H_2)$ or $N(H_2)/E_{J-K}$ or both vary 
significantly between these clouds.  Other clouds display similar
variations.  \citet{kra99} note changes in $N(CO)/A_V$ within the 
dark cloud IC 5146, which they interpret as differences in the amount 
of CO depletion onto dust grains.  From observations of C$^{17}$O, 
C$^{34}$S, and N$_2$H$^{+}$, \citet{ber01} show that the pattern of 
gas depletion in IC 5146 is consistent with models of cloud chemistry.

Our observations of $\rho$ Oph, Cha I, and the Coalsack indicate important
differences in grain chemistry among dark clouds.  Changes in the slope 
of the near-IR extinction law appear correlated with the degree of star
formation activity or the density of the cloud or both. $\rho$ Oph is
the most active star formation region of this group and it has the
regions of largest extinction. The Coalsack is the most quiescent 
cloud and has the lowest extinction.  Testing whether this correlation
is real or a fluke of small number statistics requires high quality
extinction measurements in more dark clouds.

\section{Search for Candidate Pre-Main Sequence Stars}

Figure 4 shows the $H-K$/$J-H$ diagram for our complete survey region.
The solid line corresponds to the main sequence
locus \citep{bebr88}; the dashed lines, parallel to the reddening vector
($E_{J-H}/E_{H-K}$ $=$ 2.08 $\pm$ 0.03) derived in \S3, define the reddening
band extending from the main sequence.  We detect $\sim$ 2500 stars in Globule 2
with $K$ $<$ 14.0. 

\begin{figure}
\centering
\includegraphics[width=17cm]{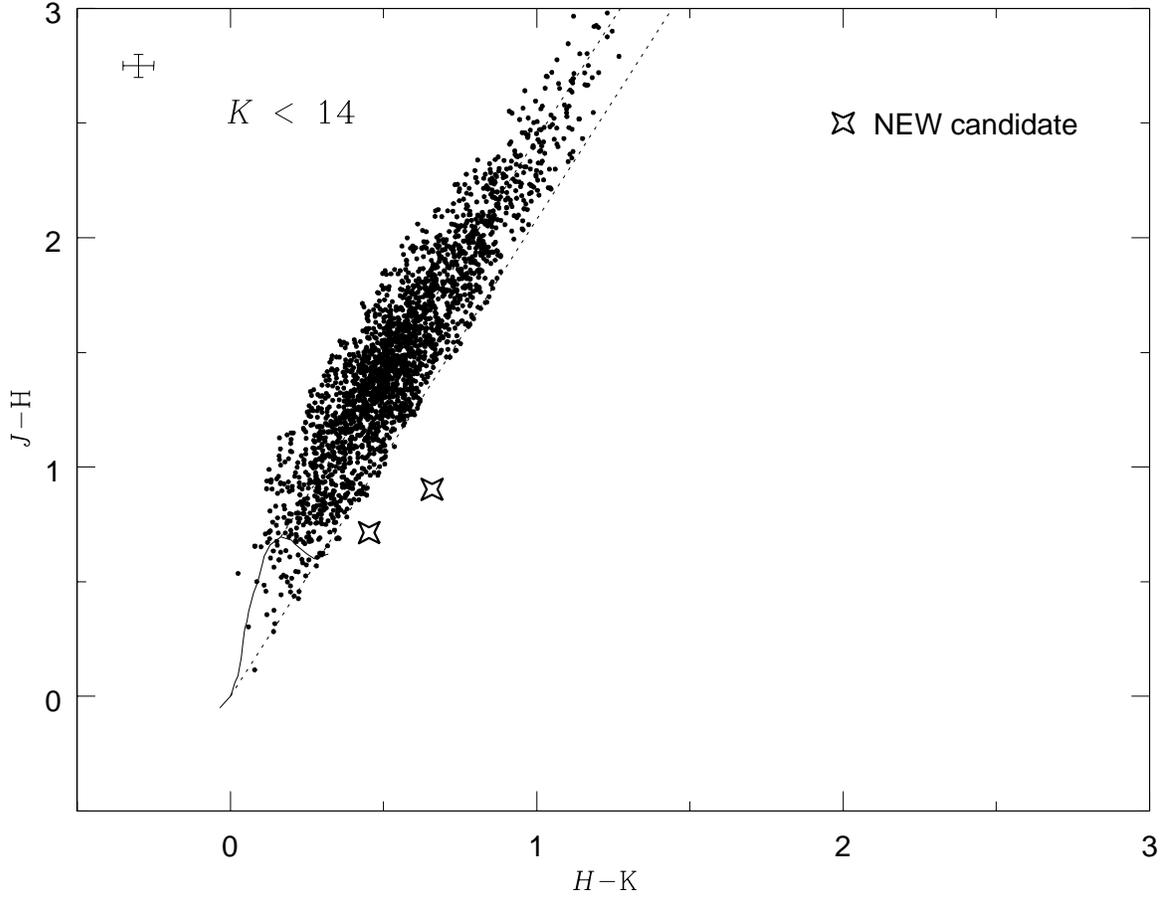}
\caption{Color-color diagram for near-IR sources
with $K$ $<$ 14.0 in our survey region. The solid line
indicates the main sequence locus \citep{bebr88} and the two parallel lines
define the reddening band, extending from the main sequence.
To define this band we use the reddening vector,
$E_{J-H}/E_{H-K}$ $=$ 2.08 $\pm$ 0.03, derived in section 3.
Typical photometric errors for $K$ $\sim$ 14.0 are displayed in the upper left
corner.  The big stars show the positions of two near-infrared candidate young
stellar objects.}
\label{Fig4}
\end{figure}                

We distinguish two groups of sources in Figure 4: i) stars that lie along
or follow the reddening band direction, and ii) two sources with 
near-IR excess located to the right of the reddening band.
The first group contains almost all of our detections. These objects 
are reddened main sequence or giant stars (see \S 3) and lie behind the
cloud \footnote{Any weak emission T Tauri stars that belong to the cloud and 
that, as a class of young stellar objects, have no significant near-IR 
excess are included in this group.}.  The color excesses of the 
second sample cannot be attributed to reddening by normal dust in 
the cloud.  Circumstellar material 
(disks + infalling envelopes) surrounding the young central stars can 
produce these color excesses at near-IR wavelengths 
\citep{laad92,keha95}. Two stars in our sample have near-IR
color excesses similar to previously known young stars 
Table 2 lists coordinates and magnitudes for these stars.
They are located within $\sim$ 5--6$'$ of the position of the 
C$^{18}$O core detected by \citet{kat99}. In a similar manner, the
nearest mid-infrared MSX sources to these candidates lie at distances of $\sim$
5$'$ and 3.5$'$, respectively. 

We find three objects in common between our near-infrared survey 
and the nine sources from the MSX catalog that lie within
radius of $\sim$ 11$'$ from the optical position of the Globule 2. 
The other 6 mid-infrared sources are located close but outside of our
15$'$ $\times$ 15$'$ observed area. 

Although the Coalsack lies in the general direction of the 
Sagittarius-Carina arm of the Milky Way, \citet{nym89} note that
there is no background cloud at the position of Globule 2. 
Confusion with background sources is not significant; 
the selected candidates probably belong to the Coalsack region lying 
in the vicinity of this globule.  The near-IR excesses of these sources
suggests the pre-main sequence nature of these stars. Optical spectroscopy
is required to test this suggestion.

\section{The Coalsack Globule 2 Density Profie}

\subsection{The Extinction Map}             

To map the distribution of extinction in Globule 2, we divide the 
observed region (a 15$'$ $\times$ 15$'$ area) into a fixed grid of 
1$'$ $\times$ 1$'$ overlapping squares, separated by the 30$''$ 
Nyquist spatial sampling interval.  We follow the same procedure 
for our off-cloud regions (see Table 1).  $H$-band star counts in
these two areas then yield an extinction map.  This standard method 
compares the stellar density in each position of the cloud with 
the density in a nearby control area assumed free of significant
obscuration. The number of stars per unit area brighter than 
m$_\lambda$ in the control field, N$_{\rm off}$, is

\begin{equation}
\rm Log~ N_{off} = a + b~ m_\lambda ~ .
\end{equation}       

On the cloud, the slope of equation (1) remains the same but the 
number of stars per unit area, N$_{\rm on}$, decreases as the extinction
Am$_\lambda$ increases. The extinction is then

\begin{equation}
\rm Am_\lambda = {1 \over b}~ Log~ {N_{off} \over N_{on}}, 
\end{equation}

\noindent
where b is the slope of the cumulative luminosity function. We use the $H$-band
luminosity function for off cloud regions to measure b $=$ 0.32 $\pm$ 0.01.
We convert the $H$-band extinction, A$_H$, to optical extinction, A$_V$, using 
the  \citet{rile85} reddening law (A$_V$ $=$ 5.7 A$_H$) and convolve the 
individual stellar measurements with a two dimensional Gaussian filter with 
a FWHM of 60$''$, the size of our grid cells. Large cells yield poor
resolution; smaller cells have too few stars for a meaningful measurement.
Figure 5 shows our extinction map. The smoothed stellar 
positions display roundish shapes. The outermost contour corresponds to 
A$_V$ $=$ 5. Subsequent
contours increase in steps of 3 mag, reaching A$_V$ $=$ 23 mag in the innermost
area. As noticed by \cite{jon84}, Globule 2 is embedded in the Coalsack;
the 5 mag level is probably the average level of dust absorption in the 
surrounding cloud region.

\begin{figure}
\centering
\includegraphics[width=17cm]{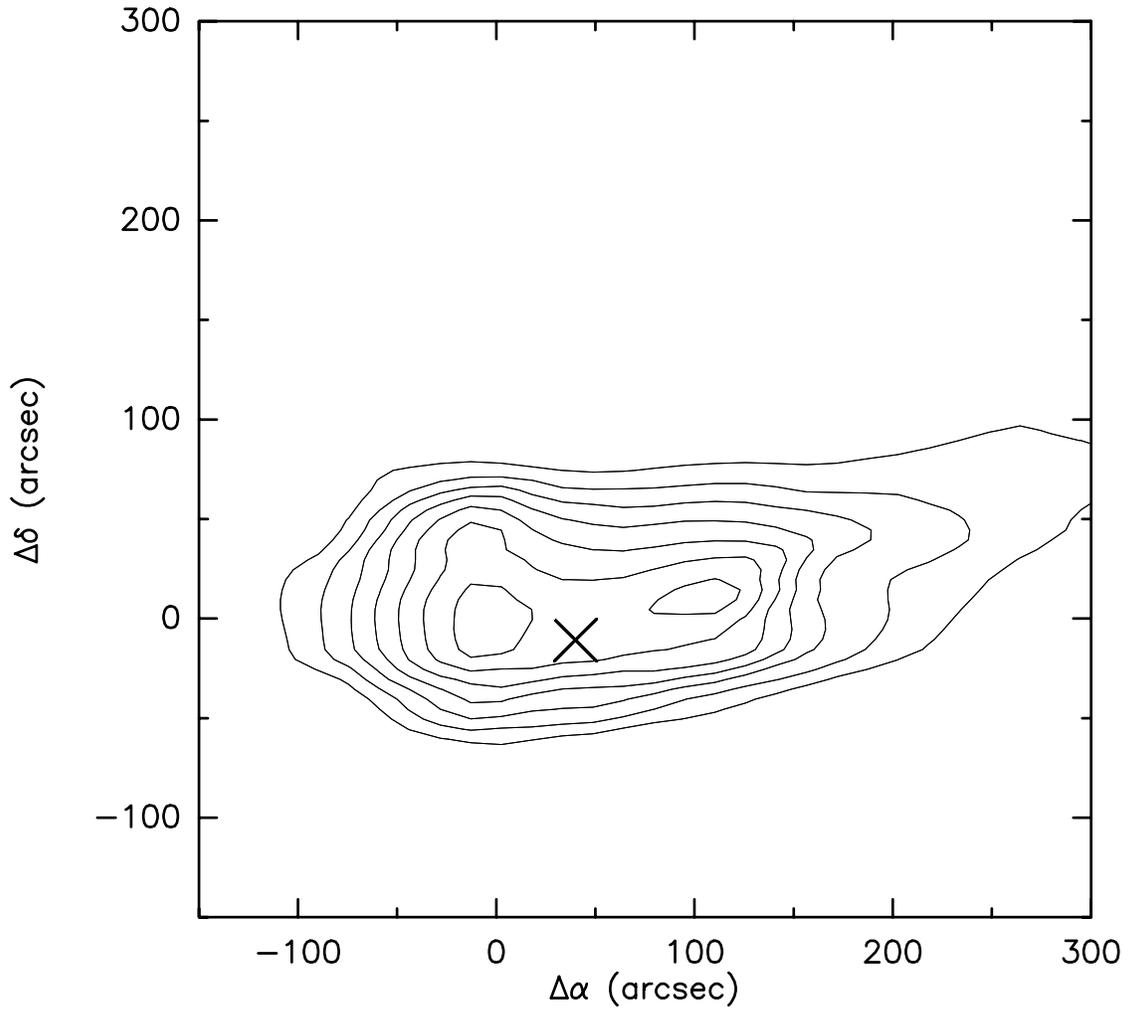}
\caption{Extinction map for the Globule 2. The lowest contour
corresponds to A$_V$ $=$ 5 mag. Subsequent contours increase in step of 3 mag.
The coordinates of the center of the region are: \hbox{$\alpha$(2000.0) $=$ 12$^h$
31$^m$ 24$^{sec}$,} \hbox{$\delta$(2000.0) $=$ $-$ 63$^o$ 45$'$ 11$''$.} The big cross
indicates the position of the \citet{kat99} C$^{18}$O core.}
\label{Fig5}
\end{figure}     

We determine mean radii for each optical obscuration level, normalize
this `average' density with respect to the 23 mag peak level, and 
plot the results in \hbox{Figure 6} as dot symbols. The projected radial dust
column density distribution is shallow, suggesting a configuration near
equilibrium.  Density profiles of unstable clouds show a clear enhancement
toward the center, indicating the collapse of material \citep[see][]{har01}.

\begin{figure}
\centering
\includegraphics[width=17cm]{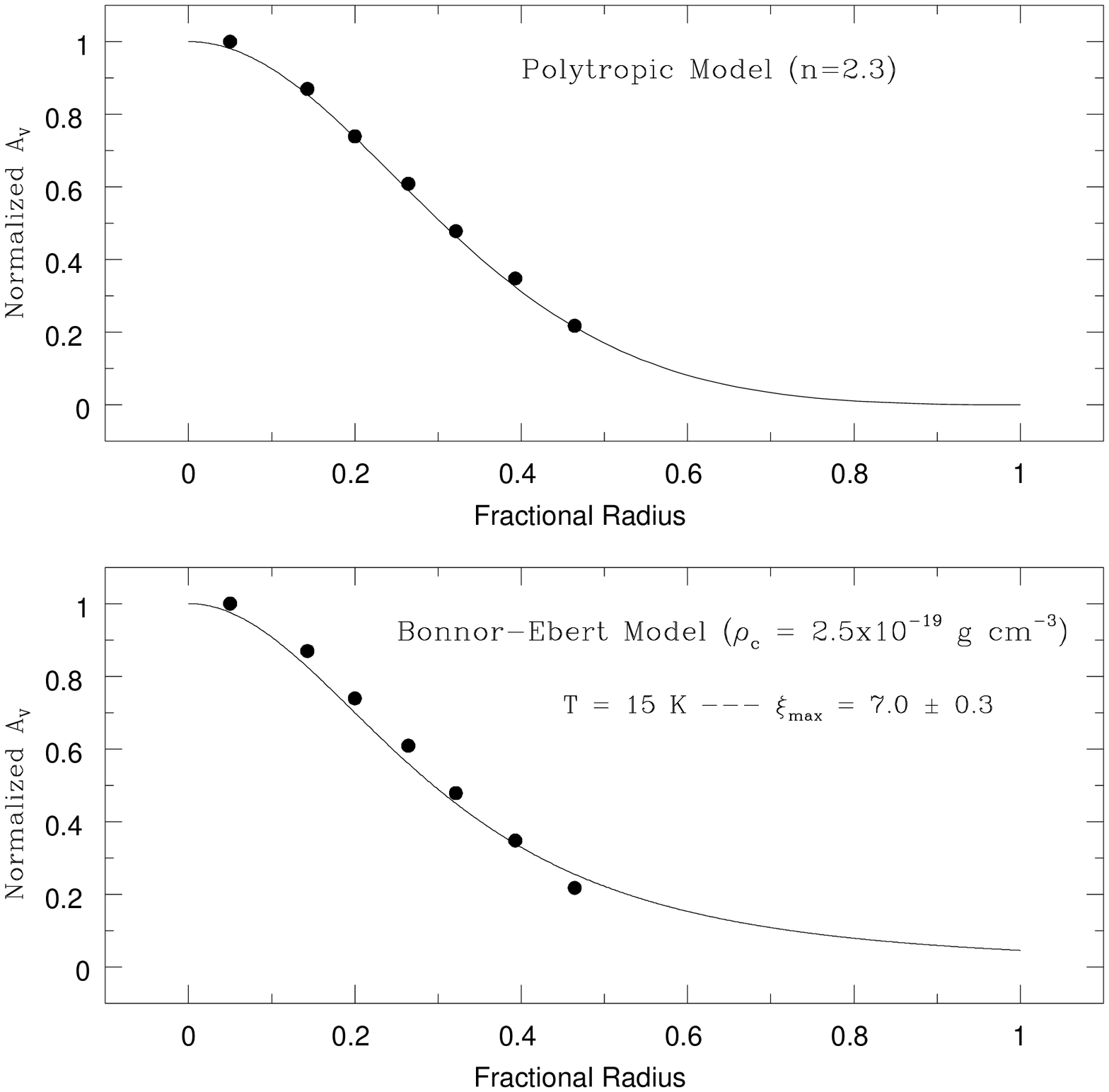}
\caption{Normalized density profile for the Globule 2. The optical
extinction A$_V$ is referred to the central peak value (23 mag). The
radial distance is the fractional radius with respect to \hbox{R $=$ 140$''$}
($\sim$ 0.1 pc), corresponding to A$_V$ $=$ 5 mag. The dot symbols represent our
observational data and the continuous lines the model fits.
The upper panel corresponds to a polytrope with \hbox{n $=$ 2.3.}
The lower panel indicates the Bonnor-Ebert
model for a central density \hbox{$\rho_c$ $=$ 2.5 $\times$ 10$^{-19}$
g/cm$^{3}$,} assuming T $=$ 15 K for this globule.}
\label{Fig6}
\end{figure}  

\subsection{The Polytropic and the Bonnor-Ebert Models}             

To model the internal structure of Globule 2, we consider the Lane-Emden 
equation for a self-gravitating isothermal gas sphere in hydrostatic 
equilibrium: 

\begin{equation}
\rm {1 \over \xi^2} {d \over d\xi} (\xi^2 {d\theta \over d\xi}) = - \theta^n, 
\end{equation}        

\noindent
where $\xi =$ r/$\alpha$ and $\theta =$ T/T$_c$ are the Emden variables, with
$\alpha$ an unspecified factor having dimensions of length and T$_c$ the
central temperature \citep[see, for example,][]{nov73}. The parameter `n'
is the polytropic index relating the gas pressure to the density, 
$P_{gas} = K_n \rho^n$.  We consider two approaches to solve equation (3): 
a) a numerical integration for specific values of `n' and 
b) the introduction of a variant of this pressure-confined, 
self-gravitating isothermal sphere equation.

\cite{kest79} adopted the first approach and
numerically solved equation (3) for different values of the
polytropic index n, satisfying the following
boundary conditions: \hbox{$\theta =$ 1} and d$\theta$/d$\xi =$ 0 at $\xi =$ 0.
The upper panel of Figure 6 shows the normalized density 
profile superposed on a polytropic model with index n $=$ 2.3. 
We adopt an `average' external radius of 140$''$, 0.1 pc
at the distance of the this cloud. This radius corresponds to A$_V =$ 5 mag,
which we assume represents the average optical extinction in the surrounding
Coalsack region. For \hbox{n $=$ 2.3.}, 
\cite{kest79}'s calculations predict a central density 
$\rho_c$ $=$ 3 $\times$ 10$^{-19}$ g/cm$^{3}$ and a temperature T $=$ 20 K 
to support the cloud against collapse.  Although we have no direct
temperature measurement of Globule 2, the average temperature for 
other small globules is T = 10--15 K \citep{bou95b}.  The factor of 1.2--2.0
difference between the `observed' and hydrostatic equilibrium temperature 
suggests that Globule 2 is marginally unstable and on the verge of collapse.

In our second approach, we consider a variant of equation (3),
known as the Bonnor-Ebert model \citep{bon56,ebe55} or 
the modified Lane-Emden equation:

\begin{equation}
\rm {1 \over \xi^2} {d \over d\xi} (\xi^2 {d\phi \over d\xi}) = e^{-\phi}, 
\end{equation} 

\noindent
where $\rho = \rho_c e^{-\phi(\xi)}$.  Here, 
$\xi = (r/a) \sqrt{ 4 \pi G \rho_c}$ is a dimensionless radial 
parameter and $a$ is the isothermal sound speed
($a  = \sqrt{kT/ \mu m_{\rm H}}$;
$\mu$ is the mean molecular weight and $m_{\rm H}$ is the mass of the H atom).
Introducing standard boundary conditions 
($\phi$(0) $=$ 0 and d$\phi$(0)/d$\xi$ $=$ 0), we solve equation (4) 
numerically. Assuming an external radius R (140$''$ or
0.1 pc), solutions of equation (4) can be parametrized by $\rho_c$.
The lower panel of Figure 6 shows our best fit 
for $\rho_c$ $=$ 2.5 $\times$ 10$^{-19}$ g/cm$^{3}$ superposed on
the Globule 2 density profile. This solution assumes T$=$ 15 K as a 
typical temperature for Bok globules \citep[see][]{bou95b} and $\mu =$ 2. 

\citet{bon56} demonstrated that the value of $\xi$ at the external radius (R), 
\hbox{$\xi_{\rm max} =$ R/$a$ $\sqrt{ 4 \pi G \rho_c}$} provides a stability measure.
Systems with $\xi_{\rm max} >$ 6.5  are unstable to gravitational collapse. 
For Globule 2, $\xi_{\rm max} =$ 7.0 $\pm$ 0.3 exceeds the critical value
and is moderately unstable.

The mass of the globule and the pressure at its outer boundary are
functions of $\xi$:

\begin{equation}
\rm M(\xi) = 4 \pi \beta^3 \rho_c (\xi^2 {d\phi \over d\xi})
\end{equation}          

\begin{equation} 
{\rm P}(\xi) = a^2 \rho_c~ {\rm e}^{-{\phi(\xi)}}, 
\end{equation}          

\noindent 
where $\beta = a /  {\rm \sqrt{4 \pi G \rho_c}}$. 
Evaluating these expressions at r $=$ R (or $\xi = \xi_{\rm max}$),
we derive \hbox{M $=$ 4.5 M$_{\odot}$} and P$_{\rm ext} =$ 9 
$\times$ 10$^{-12}$ dyn/cm$^{2}$.  \cite{kat99} estimate 10 M$_{\odot}$;
our external pressure is roughly about 6.5 times the value for the
interstellar medium \citep[see][]{elm91}.

Both the polytropic and the Bonnor-Ebert models predict a moderately
unstable configuration for Globule 2. The `typical' T $\sim$ 10-15 K 
is a factor of 2 lower than the minimum value required for hydrostatic
equilibrium.  Magnetic fields  may help to prevent the collapse.
\citet{jon84} found evidence of a mildly compressed magnetic field
from $H$-band polarization measurements of
background stars in the Globule 2 region. The very modest intensities
of magnetic fields associated with dark clouds \citep[B $\sim$ 16 $\mu$G;][]{cru93} 
suggest, however, that magnetic fields can not prevent the collapse indefinitely and
alter the global process significantly.

\citet{alv01} used the Lane-Emden equation to fit the density profile
of Barnard 68 \citep{bok77}, an isolated nearby (d $=$ 125 pc) southern
hemisphere Bok globule. They derived $\xi_{\rm max} =$ \hbox{6.9 $\pm$ 0.2,} slightly
larger than the critical value.  B68 has a radius of 0.06 pc, 
M = 2.1M$_{\odot}$, \hbox{T $=$ 16 K,} and an external pressure of~
2.5 $\times$ 10$^{-11}$ dyn/cm$^{2}$. This starless globule shows no evidence 
of star-formation and thus may be about to collapse to form stars.

\cite{har01} analyzed the dense globule in B335 \citep[d $\sim$ 250 pc;][]{tom79}. 
The shape of the B335 density profile is steeper than either Globule 2 or B68.
\cite{har01} show that collapse models and an unstable Bonnor-Ebert sphere 
(with $\xi_{\rm max} =$ 12.5 $\pm$ 2.6) match the observed density profile.
Molecular line profiles provide additional evidence that this globule is
collapsing \citep{zho93,cho95}. B335 is associated with a deeply embedded 
young stellar object \citep{kee83} that drives a bipolar outflow 
\citep{cab88,hir88,hir92}. Thus this core is in a more advanced 
evolutionary stage than B68. 

The Coalsack Globule 2 may represent an intermediate phase between B68 (a
pre-collapse globule) and B335 (an already collapsing core) according to
the polytropic models and Bonnor's critical parameter, $\xi_{\rm max}$.

\section{Summary}

Our near-IR survey of Globule 2, the highest density and more massive 
core in the Coalsack complex, leads to three primary conclusions.  

The slope of the near-IR extinction law in the cloud, $E_{J-H}/E_{H-K}$ = 
2.08 $\pm$ 0.03, is much steeper than in $\rho$ Ophiuchi or Chamaeleon I.  If real,
the trend of increasing slope with decreasing star-formation activity in 
these three clouds suggests changes in grain chemistry with extinction
or star formation activity.  

Based on near-IR excess emission, we detect two pre-main sequence 
candidates with $K < 14$ in the vicinity Globule 2.  If confirmed as pre-main sequence 
stars with optical or near-IR spectroscopy, these are the first pre-main 
sequence stars discovered in the region.  The globule contains no known 
IRAS sources \citep{nym89,bou95a} or other pre-main sequence candidates.
Our candidates are too faint for detection with IRAS.  The low success 
rate in finding pre-main sequence stars in the Globule 2 region illustrates the low 
activity of the Coalsack as stellar nursery, in agreement with previous 
investigations \citep{wea74,sch77,rei81,nym89}. 

We use $H$-band star counts to derive the density profile of this
Bok globule. For a `typical' gas temperature T $\sim$ 15 K, model
fits suggest this small cloud is moderately unstable, with a 
Bonnor critical parameter $\xi_{\rm max}$ $=$ 7.0 $\pm$ 0.3.  The
mass derived from these models, {$M = 4.5 M_{\odot}$}, agrees with
estimates derived from the CO column density. 

\acknowledgments

We are grateful to the CTIO staff, specially to M. Fern\'andez,
M. Hern\'andez, and P. Ugarte for assistance during the observing runs.
We also thank R. Elston and R. Probst for their help with CIRIM and D.
Mink for assistance with the WCSTools software. An anonymous referee
carefully read the original manuscript and provided useful suggestions.

\clearpage



\newpage





\clearpage

\newpage

\begin{deluxetable}{cccc}
\tablecaption{Off-Cloud Regions. \label{tbl-1}}
\tablewidth{0pt}
\tablehead{
\colhead{Region} & \colhead{$\alpha$(2000.0)} & \colhead{$\delta$(2000.0)} }
\startdata
I&   12 32 53& $-$64 21 34 \\
II&  12 32 55& $-$64 27 43 \\
III& 12 33 46& $-$64 26 34 \\
IV&  12 33 49& $-$64 22 44 \\
\enddata

\tablecomments{Units of right ascension are hours, minutes, and seconds, and
units of declination are degrees, arcminutes, and arcseconds.}

\end{deluxetable}

\begin{deluxetable}{cccccccl}
\tablecaption{Candidate Pre-Main Sequence Stars with $K$ $<$ 14.0.
\label{tbl-2}}
\tablewidth{0pt}
\tablehead{
\colhead{Star}& \colhead{$\alpha$(2000.0)} & \colhead{$\delta$(2000.0)} &
\colhead{$K$} &
\colhead{$H-K$} & \colhead{$J-H$} }
\startdata
1& 12 31  54.4& $-$63 50 00& 13.99&  0.66&  0.90 \\
2& 12 32  05.0& $-$63 49 30& 12.56&  0.45&  0.72 \\             
\enddata
 
\tablecomments{Units of right ascension are hours, minutes, and seconds, and
units of declination are degrees, arcminutes, and arcseconds.}
\end{deluxetable}

\end{document}